# Statistical inference methods for cumulative incidence function curves at a fixed point in time[§]

Jinbao Chen[1], Yawen Hou[2] and Zheng Chen[1*]

[1] *Department of Biostatistics, School of Public Health, Southern Medical University, Guangzhou 510515, China*; [2] *Department of Statistics, College of Economics, Jinan University, Guangzhou 510632, China*

∗: Corresponding author

**Abstract**

Competing risks data arise frequently in clinical trials. When the proportional subdistribution hazard assumption is violated or two cumulative incidence function (CIF) curves cross, rather than comparing the overall treatment effects, researchers may be interested in focusing on a comparison of clinical utility at some fixed time points. This paper extend a series of tests that are constructed based on a pseudo-value regression technique or different transformation functions for CIFs and their variances based on Gaynor's or Aalen's work, and the differences among CIFs at a given time point are compared.

# 1  Introduction

Competing risks data arise frequently in clinical trials. Subjects in such trials may fail owing to multiple causes, and failure as a result of one cause precludes the observation of failure as a result of any of the other causes. One of the most important problems encountered is the overall homogeneity of cumulative incidence function (CIF) curves (Berry et al. 2010, Iacobelli et al. 2013), Gray (1998) developed a log-rank type test is based on the subdistribution hazard (SDH), and Pepe and Mori (1993) and Bajorunaite and Klein (2007) proposed tests based on the CIF that have different variance estimators. However, these methods lose power when the assumption of proportional SDH is violated, especially when two CIF curves cross each other (Geskus 2015). The Renyi-type test (Bajorunaite and Klein 2007) and Kolmogorov-Smirnov (KS)-type test (Lin 1997, Freidlin and Korn 2005), which compare the maximum difference between SDHs and CIFs, respectively, do not rely on this assumption, but these tests exhibit only slightly improved power relative to the aforementioned methods in this situation (Bajorunaite and Klein 2007, Freidlin and Korn 2005, Bajorunaite and Klein 2006). For example, in a pediatric cancer trial study (Tai, Grundy, Machin 2010) with an ependymoma group and another brain tumor group, irradiation after disease progression was the event of interest, and declining radiotherapy (RT) and opting for elective RT were competing events. The main results are summarized in Figure 4a in Tai et al. (2010). However, because the original data are not publicly available, we reconstructed similar data (Fig. 1) via simulation. The results of Gray's test ($\chi^2 = 1.851, P = 0.174$) and Pepe and Mori's test

($\chi^2 = 0.069, P = 0.793$) indicate that the CIF of the event of interest is not significantly different between the two groups. However, the two CIF curves cross each other at an early time, near 3 years, and a goodness-of-fit test (Li, Thomas, Zhang 2015) of the proportional SDH assumption indicates that the assumption is violated ($P$=0.002). Gray's test (Gray 1998) and Pepe and Mori's test (Pepe and Mori 1993) generally fail in the case of assumption violations, with crossing CIFs or crossing SDHs, because positive differences are cancelled by negative differences, leaving the test methods unable to detect the overall differences, similar to the log-rank test for single endpoint data (Li et al. 2015, Klein and Moeschberger 2003). Moreover, the Renyi-type test (Bajorunaite and Klein 2007) ($Stat = 1.175, P = 0.479$) and the KS-type test (Lin 1997, Freidlin and Korn 2005) ($Stat = 0.221, P = 0.092$) also fail to identify a significant difference between the two groups. However, the treatment effectiveness before or after the crossing of two CIF curves might differ, as evident from Fig. 1; in particular, there is a clear difference after 3 years. The phenomenon of crossing CIF curves is common (Fine and Gray 1999) when two treatments offer different benefits before and after the crossing point. Therefore, rather than comparing entire CIF curves, investigators might be interested in comparing the (cumulative) probability of occurrence for an event of interest at a fixed time point in the presence of other risks (Lin 1997). As in the pediatric cancer trials described above, the 5-year treatment effectiveness is often used as an indicator of disease progression or prognosis (Tonorezos et al. 2015).

One approach used to compare the differences between two CIF curves at specific time points involves the construction of point-wise confidence intervals (CIs) for two CIF curves. Unfortunately, a normal-approximation CI procedure for CIF curves can result in the bounds being negative or greater than 1 when the sample size is small. However, some transformation functions can help restrict the CI coverage (Choudhury 2002, Hong and Meeher 2014). Another approach involves constructing a test statistic. Klein et al. (2007) and Su et al. (2011) proposed five test methods based on linear, logarithmic, log-log, arcsine-square-root and logit transformations to compare the difference between two independent samples for a single endpoint at time point $t$, and a method based on the pseudo-value regression technique (Klein and Andersen 2005, Klein et al. 2008, Andersen, Klein and Rosthøj 2003) was proposed to similarly compare the difference between paired or clustered samples. Lin (1997) also discussed the occurrence probability of an event of interest at some time points with competing risks; however, he did not evaluate any approaches. In addition, corresponding variance estimation for the CIF has been discussed (Braun and Yuan 2007, Allignol, Schumacher and Beyersmann 2010), and although Aalen's variance (Aalen 1978) has been programmed and documented for general use (via the *cuminc* package of R), it tends to overestimate the true variance (Braun and Yuan 2007). The approaches of Gaynor et al. (1993) and Betensky and Schoenfeld (2001), which are equivalent to one another and are based on Dinse and Larson's work (Dinse and Larson 1986) are fairly accurate, even with small samples (Braun and Yuan 2007). Thus, in the present work, several of the above-mentioned methods (Klein et al. 2007, Su et al. 2011) are extended to the analysis of competing risks data by applying the variance estimation of Gaynor et al. (1993), as suggested by Braun and Yuan (2007), and of Aalen (1978).

The article is organized as follows. In Section 2, we extend the tests described in Klein et al. (2007) and Su et al. (2011) for testing the equality of two CIFs at a fixed point in time by applying the variance estimator based on Gaynor's (Gaynor et al. 1993) and Aalen's variance. In Section 3, we investigate the powers and type I error rates of the six test methods, with variance based on the methods of Aalen (1978) and Gaynor et al. (1993), using Monte Carlo simulation. In Section 4, the six methods are illustrated using two examples. Finally, we summarize our findings and present a general discussion.

## 2  The extended two sample test methods

Let CIF $I_{rk}(t)$ be the probability of failure as a result of cause $k$ before a given time $t$ in group $r$, where we focus on only two groups, $r=1$ and 2. The CIF of cause $k$ in group $r$ is given as

$$I_{rk}(t) = P(T \leq t, D = k),$$

where $T$ and $D$ are random variables representing the time until the first observed event and the type of event, respectively. In the presence of competing risks, the CIF of cause $k$ in group $r$ can be estimated as

$$\hat{I}_{rk}(t) = \sum_{t_{rj} \leq t} \hat{S}(t_{rj-1}) \frac{d_{rkj}}{a_{rj}},$$

where $\hat{S}(t)$ is the Kaplan-Meier estimator at time $t$ in group $r$ for all causes; $t_{rj}$ denotes the $j$th ordered event time in group $r$; and $d_{rkj}/a_{rj}$ is defined as the cause-specific hazard rate, where $d_{rkj}$ is the number of events of cause $k$, and $a_{rj}$ is the risk set at time $t_j$ in group $r$. This estimator was first considered by Aalen (1978) and Aalen and Johansen (1978). The Aalen's variance (Aalen 1978) of the CIF estimator of cause $k$ in group $r$ is given by

$$\hat{V}[\hat{I}_{rk}(t)] = \sum_{t_{rj} \leq t} \left\{ [\hat{I}_{rk}(t) - \hat{I}_{rk}(t_{rj})]^2 \frac{d_{rj}}{(a_{rj}-1)(a_{rj}-d_{rj})} \right\} + \sum_{t_{rj} \leq t} \hat{S}(t_{rj-1})^2 \frac{d_{rkj}(a_{rj}-d_{rkj})}{a_{rj}^2(a_{rj}-1)}$$
$$- 2 \sum_{t_{rj} \leq t} \left\{ [\hat{I}_{rk}(t) - \hat{I}_{rk}(t_j)] \hat{S}(t_{rj-1}) \frac{d_{rkj}(a_{rj}-d_{rkj})}{a_{rj}(a_{rj}-1)(a_{rj}-d_{rj})} \right\} \quad (1)$$

where $d_{rj}$ is the number of events at time $t_{rj}$ in group $r$ for all causes. However, Aalen's variance (Aalen 1978) tends to overestimate the true variance (Braun and Yuan 2007), whereas Gaynor's variance (Gaynor et al. 1993) is fairly accurate, even with small samples. Using a Taylor series linear approximation, Gaynor's variance (Gaynor et al. 1993) equals

$$\hat{V}[\hat{I}_{rk}(t)] = \sum_{i=1}^{j} \hat{V}(\hat{I}_{rki}) + 2 \sum_{i=1}^{j-1} \sum_{i'=i+1}^{j} \text{cov}(\hat{I}_{rki}, \hat{I}_{rki'}), \quad (2)$$

where $\hat{V}(\hat{I}_{rki}) = \hat{I}_{rki}^2 \left( \frac{(a_{ri}-d_{rki})}{d_{rki}a_{ri}} + \sum_{l=1}^{i-1} \frac{d_{rl}}{a_{rl}(a_{rl}-d_{rl})} \right)$

and $\text{cov}(\hat{I}_{rki}, \hat{I}_{rki'}) = \hat{I}_{rki} \hat{I}_{rki'} \left( -\frac{1}{a_{ri}} + \sum_{l=1}^{i-1} \frac{d_{rl}}{a_{rl}(a_{rl}-d_{rl})} \right)$, $i < i'$. This formula was presented in the more general context of a semi-Markov model by Dinse and Larson (1986). To compare the difference between two CIFs at time point $t$, the null hypothesis is specified as $H_0 : I_{1k}(t) = I_{2k}(t)$, where $I_{1k}(t)$ and $I_{2k}(t)$ are the CIFs of cause $k$ at a fixed $t$ for two groups. Some transformation functions have been studied

(Klein and Moeschberger 2003). A set of suitable transformations $\phi$ of $\hat{I}_{rk}(t)$ is defined by $\phi(\hat{I}_{rk}(t))$. Applying the delta method, the estimated variances can be expressed as

$$V[\phi(\hat{I}_{rk}(t))] = \hat{V}[\hat{I}_{rk}(t)](\phi'(\hat{I}_{rk}(t)))^2 . \qquad (3)$$

Thus, classes of test statistics based on suitable transformations $\phi$ of CIFs for two independent samples are defined as follows:

$$X^2 = \frac{[\phi(\hat{I}_{1k}(t)) - \phi(\hat{I}_{2k}(t))]^2}{\hat{V}[\phi(\hat{I}_{1k}(t))] + \hat{V}[\phi(\hat{I}_{2k}(t))]} . \qquad (4)$$

## 2.1 The linear transformation test

The first test, which is based on a linear transformation of the CIFs, is extended. Let $\phi(\hat{I}_{rk}(t)) = \hat{I}_{rk}(t)$, which naively compares the values of the CIFs between two samples; the statistic defined in equation (4) for this transformation is given by

$$X_1^2 = \frac{[\hat{I}_{1k}(t) - \hat{I}_{2k}(t)]^2}{\hat{V}[\hat{I}_{1k}(t)] + \hat{V}[\hat{I}_{2k}(t)]} .$$

This simple test has an asymptotic chi-squared distribution with one degree of freedom under the null hypothesis. Here, we consider a number of transformations, all of which asymptotically yield chi-squared distributed statistics when $I_{1k}(t) = I_{2k}(t)$, and the original variance estimator is as defined in equation (1) or (2).

## 2.2 The logarithmic transformation test

The second extended test is constructed based on a logarithmic transformation of the CIFs. Let $\phi(\hat{I}_{rk}(t)) = \log(\hat{I}_{rk}(t))$. The estimated variances of equation (3) can be expressed as

$$\hat{V}(\hat{I}_{rk}(t))\frac{1}{\hat{I}_{rk}(t)^2},$$

and the statistic in equation (4) is given by

$$X_2^2 = \frac{[\log(\hat{I}_{1k}(t))-\log(\hat{I}_{2k}(t))]^2}{\dfrac{\hat{V}[\hat{I}_{1k}(t)]}{\hat{I}_{1k}(t)^2}+\dfrac{\hat{V}[\hat{I}_{2k}(t)]}{\hat{I}_{2k}(t)^2}}.$$

## 2.3 The log-log transformation test

The third test is constructed based on the log(-log(·)) transformation of the CIF. Let $\phi(\hat{I}_{rk}(t)) = \log(-\log(\hat{I}_{rk}(t)))$. The estimated variances of formula (3) can be expressed as

$$\hat{V}(\hat{I}_{rk}(t))\frac{1}{\hat{I}_{rk}(t)^2 \log(\hat{I}_{rk}(t))^2},$$

and the statistic in formula (4) is given by

$$X_3^2 = \frac{[\log(-\log(I_{1k}(t)))-\log(-\log(I_{2k}(t)))]^2}{(\dfrac{V[I_{1k}(t)]}{I_{1k}(t)^2})\Big/\log(I_{1k}(t))^2 + (\dfrac{V[I_{2k}(t)]}{I_{2k}(t)^2})\Big/\log(I_{2k}(t))^2}.$$

## 2.4 The arcsine-square-root transformation test

The fourth test is constructed based on an arcsine-square root transformation. Let $\phi(\hat{I}_{rk}(t)) = \arcsin(\sqrt{\hat{I}_{rk}(t)})$. The estimated variances defined by equation (3) can be expressed as

$$0.25\hat{V}(\hat{I}_{rk}(t))\frac{1}{\hat{I}_{rk}(t)(1-\hat{I}_{rk}(t))},$$

and the statistic in equation (4) is given by

$$X_4^2 = \frac{[\arcsin(\hat{I}_{1k}(t)) - \arcsin(\hat{I}_{2k}(t))]^2}{\dfrac{\hat{V}[\hat{I}_{1k}(t)]/\hat{I}_{1k}(t)}{4(1-\hat{I}_{1k}(t)])} + \dfrac{\hat{V}[\hat{I}_{2k}(t)]/\hat{I}_{2k}(t)}{4(1-\hat{I}_{2k}(t)])}}.$$

## 2.5 The logit transformation test

The fifth test is based on the logit transformation. Let $\phi(\hat{I}_{rk}(t)) = \log(\hat{I}_{rk}(t)/(1-\hat{I}_{rk}(t)))$. The estimated variances defined by equation (3) can be expressed as

$$\hat{V}(\hat{I}_{rk}(t))\frac{1}{\hat{I}_{rk}(t)^2(1-\hat{I}_{rk}(t))^2},$$

and equation (4) yields

$$X_5^2 = \frac{[\log(\frac{\hat{I}_{1k}(t)}{1-\hat{I}_{1k}(t)}) - \log(\frac{\hat{I}_{2k}(t)}{1-\hat{I}_{2k}(t)})]^2}{(\frac{\hat{V}[\hat{I}_{1k}(t)]}{\hat{I}_{1k}(t)^2})\Big/(1-\hat{I}_{1k}(t))^2 + (\frac{\hat{V}[\hat{I}_{2k}(t)]}{\hat{I}_{2k}(t)^2})\Big/(1-\hat{I}_{2k}(t))^2}.$$

## 2.6 The pseudo-value regression test

The last test is a method based on a pseudo-value regression technique (Klein and Andersen 2005, Klein and Gerster 2008, Andersen, Klein and Rosthøj 2003). We fix a set of time points $\tau_h$, where $h=1, ..., m$. At each time point, we compute the pooled sample CIF, $\hat{C}(\tau_h)$, based on the entire sample of size $N$ for two groups and the estimated CIF based on the sample of size $N-1$ with the $i$th observation removed for two groups, $\hat{C}^{(i)}(\tau_h)$. Then, we define the $i$th pseudo-value at time $\tau_h$ as

$$\hat{\theta}_{ih} = N\hat{C}(\tau_h) - (N-1)\hat{C}^{(i)}(\tau_h), i=1,...,N, h=1,...,m$$

when there is no censoring, where $N\hat{C}(\tau_h)$ is the number of events with cause $k$

occurring prior to $\tau_h$. In this situation,

$$\hat{\theta}_i = (\hat{\theta}_{ih}, h=1,...,m) = (I(t_i \leq \tau_1, \delta_i = k),...,I(t_i \leq \tau_m, \delta_i = k)),$$

where the $\hat{\theta}_i$s are independent. When there is censoring, the pseudo-values are close to the indicators and are approximately independent. Pseudo-values can be used in generalized linear models to model the effects of covariates on an outcome (Andersen PK, Klein JP, Rosthøj 2003). Let $g(\cdot)$ be a link function. Possible choices of the link function in models for the cumulative incidence are the logit link, the complementary log-log function on $x$ and the complementary log-log function on $1-x$ (Klein and Andersen 2005). The following generalized linear model is assumed:

$$g(\theta_{ih}) = \beta^T Z_{ih}, \quad i=1,...,N, h=1,...,m,$$

where $\beta^T$ stands for the transpose of $\beta$, and $Z_{ih}$ is a vector of covariates. For the two-sample problem, we let $Z_{ih1} = 1$ for group 1 and $Z_{ih2} = 0$ for group 2. The inverse link is defined as

$$\theta_{ih} = g^{-1}(\beta^T Z_{ih}) = \mu(\beta^T Z_{ih}).$$

We use the generalized estimating-equation approach (GEE) of Liang and Zeger (1986) to estimate $\beta$. Let $\hat{\theta}_i = (\hat{\theta}_{i1},...,\hat{\theta}_{im})^T$ and $c_i = (C(\tau_1|Z_i),...,C(\tau_m|Z_i))^T$. Define $d\mu_i(\beta)$ to be the $(m+p) \times m$ matrix of partial derivatives of $\mu(\beta^T Z_i)$ with respect to the parameters. Let $V_i(\beta)$ be a working covariance matrix. The estimating equations to be solved are then

$$U(\beta) = \sum_i d\mu_i(\beta) V_i^{-1}(\beta)(\hat{\theta}_i - c_i) = \sum_i U_i(\beta) = 0.$$

Let $\beta$ be the solution to this system of equations, and note that using results from Liang and Zeger (1986), under standard regularity conditions, we can see that

$n^{1/2}(\hat{\beta} - \beta)$ is asymptotically normal with mean zero and a covariance that can be estimated consistently using the "sandwich" estimator given by

$$\hat{\Sigma} = D(\hat{\beta})^{-1}\hat{V}(U(\hat{\beta}))D(\hat{\beta})^{-1},$$

where $D(\beta) = \sum_i d\mu_i(\beta)V_i(\beta)^{-1}d\mu_i(\beta)^T$ and $\hat{V}(U(\beta)) = \sum_i U_i(\beta)U_i(\beta)^T$. There are three possible suggestions for the working covariance matrix (Klein and Andersen 2005). Therefore, a test of the equality at time *t* of the CIFs between two samples based on $\hat{\beta}$ and covariance $\hat{\Sigma}$ can be proposed, with the hypothesis $H_0 : \beta_2 = 0$ at a fixed time. The link functions we focus on are the logit link and the log-log function on *x*.

## 3  Simulations

A Monte Carlo simulation study was designed to evaluate the performance of the test methods in terms of type I error and power. We used the terms Linear, Log, Llog, Arcs, and Logit to refer to the tests of the linear, logarithmic, log-log, arcsine-square-root and logit transformations based on Gaynor's variance (Gaynor et al. 1993) or Aalen's variance (Aalen 1978). We considered equal sample sizes ($n_1=n_2=50$, 100, 200) and unequal sample sizes ($n_1=50$, $n_2=100$; $n_1=100$, $n_2=200$) in each group. The censoring times for the two groups were generated from uniform distributions, and the overall censoring fraction in either setup was fixed at 0%, 15%, 30% or 45%. All tests with nominal level $\alpha = 0.05$ were applied to each sample, and all simulations were performed using 10000 replications.

    The failure times of an event of interest were generated from the CIF:

$I_1(t) = 1 - [1 - p(1 - e^{-t})]^{\exp(\beta Z)}$, where $p = 0.66$; thus, the maximum cumulative incidence of the event was set to 66%. $Z$ was used as the group indicator (group 1: $Z = 0$, group 2: $Z = 1$). The failure times of the competing event were generated from the CIF: $I_2(t) = (1-p)^{\exp(\beta Z)}(1 - e^{-t\exp(\beta Z)})$. The subdistribution hazard ratio (SHR) equals $\exp(\beta)$, which means that the SDH of the event for group 2 is $\exp(\beta)$ times the SDH for group 1. The type I error rate was evaluated under the null hypothesis by setting SHR: $\exp(\beta) = 1$, and failure times for both groups were generated from the same CIFs for the event of interest and the competing event. Power was evaluated for the two different scenarios by setting SHR=1.5 or 2. The fixed time was set as $t = 0.5$ or 1.

Table 1 shows the empirical type I error rates. All extended tests preserve reasonable type I error rates as the sample size becomes large ($n_1$>50, $n_2$>50). The performances of the log-log transformation test based on Gaynor's variance (Gaynor et al. 1993) and arcsine-square-root transformation test based on Aalen's variance (Aalen 1978) are much better (type I error rates close to 0.05) than the performances of the other tests. However, Aalen's variance (Aalen 1978) tends to overestimate the true variance (Braun and Yuan 2007), whereas Gaynor's variance (1993) is fairly accurate. Table 2 presents the power of the tests based on Gaynor's variance (Gaynor et al. 1993). The power of the tests increases as the sample size increases, the SHR increases, or the fixed time point increases. In contrast, the power of the tests decreases as the censoring proportion increases.

To summarize the considerable simulation results from table 1 and table 2, we applied analysis of variance (ANOVA) techniques (Klein et al. 2007, Su et al. 2011) to evaluate both the type I error and power. To evaluate the type 1 error rate, the response variable, *Y*, was defined as the percent rejection rate minus the nominal 5% level. In this way, good performance of the test is implied by absolute small, close-to-zero estimates for the expectation *E*(*Y*) in the ANOVA. To evaluate power, the outcome variable, *Y*, was defined as the percent rejection rate; good performance is indicated by large estimated values of *E*(*Y*). Here we considered four different factors: *TEST*, with 12 levels; $NUM_1\_NUM_2$, with 5 levels; *TIME*, with 2 levels; and *CEN*, with 4 levels. These factors represent the test method, sample size of each group, fixed time point and censoring proportion, respectively. We fit four models for *E*(*Y*) without intercepts as follows:

Model 1: $E(Y) = TEST \times NUM_1\_NUM_2 + TIME + CEN$,

Model 2: $E(Y) = TEST \times TIME + NUM_1\_NUM_2 + CEN$,

Model 3: $E(Y) = TEST \times CEN + TIME + NUM_1\_NUM_2$, and

Model 4: $E(Y) = TEST + CEN + TIME + NUM_1\_NUM_2$.

For example, if we focus on the situations with equal and unequal sample sizes, we can fit model 1 to evaluate the factor $TEST \times NUM_1\_NUM_2$ adjusted for the effects of the other two factors. To illustrate the performance of tests according to the factors fixed time point and censoring proportion, we can fit Models 2 and 3. For comparison, we also fit the additive Model 4.

Table 3 lists the average deviations from the nominal 5% level of the twelve proposed tests through fitted Models 1-4, and the last row presents the marginal effects of *TEST* from Model 4. We see that the tests based on the linear and arcsine-square-root transformations tend to have slightly elevated type I error rates, whereas the other tests are slightly conservative. The log-log transformation test based on Gaynor's variance (Gaynor et al. 1993) and the arcsine-square-root transformation test based on Aalen's variance (Aalen 1978) perform the best, with average deviations from the nominal 5% level close to 0. However, Aalen's variance (Aalen 1978) tents to overestimate the true variance (Braun and Yuan 2007), whereas Gaynor's variance (Gaynor et al. 1993) is fairly accurate. The average deviations of the Logit test based on pseudo-values were less than those of the log-log test based on pseudo-values. Table 4 presents the average rejection rates for the twelve tests using ANOVA with Models 1-4. Power increases with increasing sample size, increasing SHR, increasing fixed time point and decreasing censoring proportion. The log transformation tests have lower power. Although the linear transformation and arcsine-square-root transformation tests have higher power, they are anti-conservative. The power of the Logit test based on pseudo-values was greater than that of the log-log test based on pseudo-values. In summary, the simulation suggests that log-log transformation based on Gaynor's variance (Gaynor et al. 1993) is a satisfactory method for comparing competing risks data at a fixed time point.

## 4 Examples

Example 1: Tai et al. (2010) described data from trials conducted from December 1992 to April 2003 that were designed to delay or avoid irradiation of children with malignant brain tumors. The original data are not publicly available; thus, we generated similar data using a simulation method (Royston and Parmar 2011). After pathology review, the following diagnostic groups were included: ependymoma ($n=104$) and other brain tumors ($n=75$), the latter of which included astrocytoma, medulloblastoma, choroid-plexus carcinoma and mixed glioma. Each group had two causes: the event of interest was irradiation after disease progression, and the competing event was declining radiotherapy (RT) or opting for elective RT. For ependymoma, there were 48 patients for the event of interest and 31 patients for the competing event. For other brain tumors, there were 26 patients for the event of interest and 44 patients for the competing event. However, the proportional SDH assumption was violated ($P=0.002$). As presented in table 5, Gray's test, Pepe and Mori's test, the Renyi test and the KS test for the overall hypothesis of CIFs all indicated that there was no significant difference in the CIF for the interesting event between the two groups ($P>0.05$). However, comparing the CIFs at 1, 3, 5 and 7 years between the two groups, the log-log transformation test based on Gaynor's variance (Gaynor et al. 1993) indicated that there were significant statistical differences ($P<0.05$) at 1, 5 and 7 years but not at 3 years. The CIFs of the other brain tumors group were greater than those of the ependymoma group at 1 year; conversely, the differences between the two groups had opposite signs at 5 and 7 years.

Example 2: Data were obtained from the European Group for Blood and Marrow Transplantation (EBMT) registry on 2279 acute lymphoid leukemia patients who had an allogeneic bone marrow transplant from an HLA-identical sibling donor between

1985 and 1998 (de Wreede, Fiocco and Putter 2011). For the purposes of the present study, we defined the event of interest as death from transplantation, and the competing event was relapse from transplantation. Group one comprised patients with donor-recipient gender mismatch ($n$=1734), and group two comprised patients with no gender mismatch ($n$=545). For group one, there were 90 patients for the event of interest, and 145 patients for the competing event. For group two, there were 388 patients for the event of interest, and 280 patients for the competing event. However, the proportional SDH assumption was violated ($P$=0.001). Fig. 2 shows that the two CIF curves close nearly at an early time point and that there is a clear difference at a later time. Gray's test, Pepe and Mori's test and the Renyi test for the overall hypothesis of the CIFs all indicated no significant difference ($P>0.05$) for the event of interest between the two groups (table 6), whereas the KS test indicated a significant difference. In comparing the CIFs at 1000, 2000, 3000, 4000 and 5000 days between the two groups, all methods indicated significant statistical differences ($P<0.05$) at 4000 and 5000 days but not at 1000, 2000 or 3000 days. In summary, the CIFs of the gender-mismatch group were greater than those of the no-gender-mismatch group at 4000 and 5000 days but not at 1000, 2000 or 3000 days.

## 5   Concluding remarks

In this paper, we extended a series of transformation test methods based on Gaynor's (1993) or Aalen's variance (Aalen 1978) at a fixed time point for competing risks data. When the proportional SDH assumption is violated, in particular by the crossing of two CIF curves, Gray's test and Pepe and Mori's test for overall CIFs are not reliable for comparing the overall homogeneity. In this situation, researchers might be interested in comparing the differences at specific time points rather than comparing the overall hypotheses of the CIFs. ANOVA of the Monte Carlo simulation results confirmed that the log-log transformation test based on Gaynor's variance (Gaynor et al. 1993) had robust power under various situations, with reasonable type I error rates. In each of the examples presented in this paper, the two CIF curves crossed each other, and the overall homogeneity tests revealed no significant difference between the two groups. However, the point tests for comparing two CIFs at a fixed time yielded appropriate results at the same fixed time points.

Faced with the problem of comparing the CIFs of more than two sample groups at a fixed time, the extension of the test methods to the case of $R>2$ groups might be of interest. Tests based on different transformations of CIFs can be defined in a similar manner and can be constructed using a quadratic form $X^2 = A\Sigma^{-1}A^T$, where $A$ is the vector $[\phi(\hat{I}_{1k}(t)) - \phi(\hat{I}_{2k}(t)),..., \phi(\hat{I}_{1k}(t)) - \phi(\hat{I}_{Rk}(t))]$ and $\Sigma$ is the $(R-1)\times(R-1)$ matrix with diagonal elements $[\phi(\hat{I}_{1k}(t)) - \phi(\hat{I}_{2k}(t)),..., \phi(\hat{I}_{1k}(t)) - \phi(\hat{I}_{Rk}(t))]$ and off-diagonal elements $V[\phi(\hat{I}_{1k}(t))] + V[\phi(\hat{I}_{rk}(t))]$. The pseudo-value technique can be generalized to the $R>2$ group situation using $R$-1 indicator variables or dummy variables.

**Funding**

This work is supported by the National Natural Science Foundation of China (81673268) and Natural Science Foundation of Guangdong Province (2017A030313812).

**Figure captions:**

Fig. 1　Cumulative incidence of progression/irradiation for ependymoma (solid line) and other brain tumor (dashed line) for the event irradiation after disease progression

Fig. 2　Cumulative incidence of death/relapse for no gender mismatch (solid line) and gender mismatch (dashed line) for the event death from transplantation

**Table 1** Empirical type I error rates of several tests for competing risks data

| t point | $n_1$ | $n_2$ | cens | Gaynor | | | | | Aalen | | | | | Pseudo | |
|---|---|---|---|---|---|---|---|---|---|---|---|---|---|---|---|
| | | | | Linear | Log | Llog | Arcs | Logit | Linear | Log | Llog | Arcs | Logit | Llog | Logit |
| 0.5 | 50 | 50 | 0 | 0.055 | 0.039 | 0.051 | 0.054 | 0.046 | 0.053 | 0.042 | 0.050 | 0.053 | 0.048 | 0.048 | 0.053 |
| | | | 0.15 | 0.055 | 0.039 | 0.050 | 0.053 | 0.045 | 0.053 | 0.039 | 0.049 | 0.052 | 0.045 | 0.047 | 0.049 |
| | | | 0.30 | 0.052 | 0.038 | 0.047 | 0.051 | 0.043 | 0.051 | 0.038 | 0.047 | 0.049 | 0.042 | 0.044 | 0.046 |
| | | | 0.45 | 0.051 | 0.036 | 0.048 | 0.050 | 0.043 | 0.050 | 0.036 | 0.047 | 0.049 | 0.042 | 0.043 | 0.045 |
| | 150 | 150 | 0 | 0.050 | 0.046 | 0.049 | 0.050 | 0.048 | 0.050 | 0.046 | 0.048 | 0.050 | 0.048 | 0.048 | 0.050 |
| | | | 0.15 | 0.051 | 0.047 | 0.050 | 0.051 | 0.048 | 0.051 | 0.046 | 0.049 | 0.050 | 0.048 | 0.048 | 0.049 |
| | | | 0.30 | 0.050 | 0.046 | 0.049 | 0.050 | 0.048 | 0.050 | 0.045 | 0.049 | 0.050 | 0.047 | 0.048 | 0.049 |
| | | | 0.45 | 0.049 | 0.044 | 0.048 | 0.049 | 0.047 | 0.049 | 0.044 | 0.047 | 0.048 | 0.046 | 0.046 | 0.047 |
| | 200 | 200 | 0 | 0.050 | 0.047 | 0.049 | 0.050 | 0.048 | 0.049 | 0.047 | 0.049 | 0.049 | 0.048 | 0.048 | 0.049 |
| | | | 0.15 | 0.052 | 0.048 | 0.050 | 0.051 | 0.049 | 0.052 | 0.048 | 0.050 | 0.051 | 0.049 | 0.050 | 0.050 |
| | | | 0.30 | 0.050 | 0.047 | 0.049 | 0.049 | 0.048 | 0.050 | 0.047 | 0.048 | 0.049 | 0.047 | 0.048 | 0.048 |
| | | | 0.45 | 0.049 | 0.045 | 0.048 | 0.048 | 0.047 | 0.048 | 0.045 | 0.047 | 0.048 | 0.046 | 0.047 | 0.047 |
| | 50 | 100 | 0 | 0.050 | 0.040 | 0.046 | 0.049 | 0.044 | 0.049 | 0.040 | 0.047 | 0.048 | 0.044 | 0.045 | 0.045 |
| | | | 0.15 | 0.055 | 0.042 | 0.049 | 0.052 | 0.045 | 0.053 | 0.041 | 0.048 | 0.051 | 0.044 | 0.044 | 0.046 |
| | | | 0.30 | 0.054 | 0.039 | 0.049 | 0.051 | 0.044 | 0.052 | 0.039 | 0.047 | 0.050 | 0.042 | 0.043 | 0.044 |
| | | | 0.45 | 0.053 | 0.040 | 0.048 | 0.051 | 0.043 | 0.052 | 0.040 | 0.046 | 0.050 | 0.042 | 0.043 | 0.044 |
| | 100 | 200 | 0 | 0.055 | 0.047 | 0.051 | 0.052 | 0.050 | 0.055 | 0.047 | 0.052 | 0.052 | 0.049 | 0.051 | 0.050 |
| | | | 0.15 | 0.055 | 0.047 | 0.052 | 0.053 | 0.049 | 0.055 | 0.047 | 0.050 | 0.052 | 0.048 | 0.049 | 0.049 |
| | | | 0.30 | 0.054 | 0.049 | 0.052 | 0.053 | 0.050 | 0.054 | 0.048 | 0.051 | 0.053 | 0.050 | 0.050 | 0.051 |
| | | | 0.45 | 0.053 | 0.047 | 0.050 | 0.052 | 0.049 | 0.052 | 0.046 | 0.050 | 0.051 | 0.047 | 0.049 | 0.049 |
| 1 | 50 | 50 | 0 | 0.055 | 0.044 | 0.055 | 0.055 | 0.055 | 0.055 | 0.039 | 0.050 | 0.055 | 0.050 | 0.055 | 0.055 |
| | | | 0.15 | 0.057 | 0.044 | 0.051 | 0.054 | 0.051 | 0.053 | 0.042 | 0.047 | 0.051 | 0.046 | 0.051 | 0.052 |
| | | | 0.30 | 0.056 | 0.044 | 0.050 | 0.053 | 0.049 | 0.053 | 0.040 | 0.046 | 0.050 | 0.046 | 0.050 | 0.051 |
| | | | 0.45 | 0.059 | 0.042 | 0.049 | 0.054 | 0.049 | 0.052 | 0.040 | 0.044 | 0.049 | 0.045 | 0.048 | 0.048 |
| | 150 | 150 | 0 | 0.054 | 0.051 | 0.053 | 0.054 | 0.053 | 0.054 | 0.048 | 0.051 | 0.053 | 0.051 | 0.054 | 0.054 |
| | | | 0.15 | 0.051 | 0.048 | 0.050 | 0.050 | 0.050 | 0.050 | 0.047 | 0.049 | 0.050 | 0.049 | 0.050 | 0.050 |
| | | | 0.30 | 0.049 | 0.044 | 0.046 | 0.048 | 0.046 | 0.047 | 0.042 | 0.045 | 0.046 | 0.045 | 0.046 | 0.046 |
| | | | 0.45 | 0.050 | 0.046 | 0.048 | 0.049 | 0.048 | 0.049 | 0.044 | 0.046 | 0.048 | 0.046 | 0.047 | 0.048 |
| | 200 | 200 | 0 | 0.049 | 0.049 | 0.049 | 0.049 | 0.049 | 0.049 | 0.049 | 0.049 | 0.049 | 0.049 | 0.049 | 0.049 |
| | | | 0.15 | 0.052 | 0.048 | 0.050 | 0.051 | 0.050 | 0.050 | 0.047 | 0.049 | 0.050 | 0.049 | 0.050 | 0.050 |
| | | | 0.30 | 0.051 | 0.048 | 0.050 | 0.050 | 0.050 | 0.050 | 0.048 | 0.049 | 0.050 | 0.049 | 0.050 | 0.050 |
| | | | 0.45 | 0.054 | 0.050 | 0.051 | 0.053 | 0.051 | 0.052 | 0.049 | 0.050 | 0.051 | 0.050 | 0.051 | 0.051 |
| | 50 | 100 | 0 | 0.055 | 0.043 | 0.048 | 0.053 | 0.049 | 0.051 | 0.042 | 0.046 | 0.048 | 0.043 | 0.047 | 0.049 |
| | | | 0.15 | 0.055 | 0.047 | 0.048 | 0.051 | 0.048 | 0.051 | 0.043 | 0.045 | 0.048 | 0.045 | 0.049 | 0.048 |
| | | | 0.30 | 0.054 | 0.046 | 0.048 | 0.051 | 0.048 | 0.051 | 0.042 | 0.045 | 0.048 | 0.044 | 0.046 | 0.047 |
| | | | 0.45 | 0.057 | 0.045 | 0.049 | 0.054 | 0.048 | 0.053 | 0.041 | 0.045 | 0.050 | 0.043 | 0.046 | 0.045 |
| | 100 | 200 | 0 | 0.056 | 0.049 | 0.051 | 0.054 | 0.053 | 0.054 | 0.048 | 0.050 | 0.052 | 0.050 | 0.050 | 0.053 |
| | | | 0.15 | 0.055 | 0.050 | 0.052 | 0.053 | 0.051 | 0.053 | 0.049 | 0.050 | 0.050 | 0.049 | 0.051 | 0.051 |
| | | | 0.30 | 0.056 | 0.051 | 0.052 | 0.054 | 0.052 | 0.054 | 0.049 | 0.051 | 0.052 | 0.050 | 0.052 | 0.052 |
| | | | 0.45 | 0.061 | 0.055 | 0.055 | 0.059 | 0.057 | 0.058 | 0.053 | 0.053 | 0.056 | 0.054 | 0.055 | 0.055 |

**Table 2** Empirical power of several tests based on Gaynor's variance for competing risks data

| t point | $n_1$ | $n_2$ | cens | SHR=1.5 | | | | | SHR=2 | | | | |
|---|---|---|---|---|---|---|---|---|---|---|---|---|---|
| | | | | Linear | Log | Llog | Arcs | Logit | Linear | Log | Llog | Arcs | Logit |
| 0.5 | 50 | 50 | 0 | 0.407 | 0.381 | 0.398 | 0.403 | 0.392 | 0.882 | 0.863 | 0.881 | 0.881 | 0.881 |
| | | | 0.15 | 0.402 | 0.359 | 0.387 | 0.395 | 0.381 | 0.863 | 0.838 | 0.850 | 0.858 | 0.849 |
| | | | 0.30 | 0.387 | 0.345 | 0.369 | 0.380 | 0.364 | 0.834 | 0.808 | 0.817 | 0.826 | 0.817 |
| | | | 0.45 | 0.359 | 0.319 | 0.341 | 0.352 | 0.338 | 0.782 | 0.760 | 0.759 | 0.775 | 0.765 |
| | 150 | 150 | 0 | 0.852 | 0.844 | 0.850 | 0.851 | 0.849 | 1.000 | 1.000 | 1.000 | 1.000 | 1.000 |
| | | | 0.15 | 0.835 | 0.825 | 0.830 | 0.833 | 0.829 | 0.999 | 0.999 | 0.999 | 0.999 | 0.999 |
| | | | 0.30 | 0.810 | 0.801 | 0.805 | 0.808 | 0.804 | 0.999 | 0.999 | 0.999 | 0.999 | 0.999 |
| | | | 0.45 | 0.770 | 0.763 | 0.766 | 0.769 | 0.766 | 0.997 | 0.997 | 0.997 | 0.997 | 0.997 |
| | 200 | 200 | 0 | 0.937 | 0.933 | 0.936 | 0.936 | 0.935 | 1.000 | 1.000 | 1.000 | 1.000 | 1.000 |
| | | | 0.15 | 0.924 | 0.921 | 0.922 | 0.923 | 0.922 | 1.000 | 1.000 | 1.000 | 1.000 | 1.000 |
| | | | 0.30 | 0.911 | 0.908 | 0.909 | 0.911 | 0.909 | 1.000 | 1.000 | 1.000 | 1.000 | 1.000 |
| | | | 0.45 | 0.883 | 0.880 | 0.881 | 0.883 | 0.882 | 1.000 | 1.000 | 1.000 | 1.000 | 1.000 |
| | 50 | 100 | 0 | 0.545 | 0.445 | 0.526 | 0.524 | 0.500 | 0.957 | 0.926 | 0.956 | 0.956 | 0.948 |
| | | | 0.15 | 0.527 | 0.427 | 0.509 | 0.506 | 0.475 | 0.946 | 0.915 | 0.944 | 0.942 | 0.936 |
| | | | 0.30 | 0.504 | 0.401 | 0.485 | 0.482 | 0.453 | 0.932 | 0.894 | 0.929 | 0.927 | 0.919 |
| | | | 0.45 | 0.471 | 0.363 | 0.449 | 0.448 | 0.417 | 0.897 | 0.857 | 0.892 | 0.890 | 0.881 |
| | 100 | 200 | 0 | 0.814 | 0.774 | 0.810 | 0.808 | 0.798 | 0.999 | 0.999 | 0.999 | 0.999 | 0.999 |
| | | | 0.15 | 0.801 | 0.755 | 0.795 | 0.790 | 0.777 | 0.999 | 0.998 | 0.999 | 0.999 | 0.999 |
| | | | 0.30 | 0.776 | 0.728 | 0.768 | 0.765 | 0.752 | 0.998 | 0.997 | 0.998 | 0.998 | 0.998 |
| | | | 0.45 | 0.741 | 0.692 | 0.733 | 0.730 | 0.719 | 0.996 | 0.994 | 0.996 | 0.996 | 0.995 |
| 1 | 50 | 50 | 0 | 0.572 | 0.519 | 0.572 | 0.572 | 0.572 | 0.955 | 0.952 | 0.954 | 0.955 | 0.955 |
| | | | 0.15 | 0.524 | 0.489 | 0.498 | 0.514 | 0.501 | 0.924 | 0.913 | 0.912 | 0.921 | 0.915 |
| | | | 0.30 | 0.474 | 0.435 | 0.438 | 0.460 | 0.445 | 0.809 | 0.795 | 0.781 | 0.801 | 0.789 |
| | | | 0.45 | 0.327 | 0.301 | 0.277 | 0.311 | 0.291 | 0.683 | 0.674 | 0.600 | 0.660 | 0.631 |
| | 150 | 150 | 0 | 0.950 | 0.940 | 0.945 | 0.950 | 0.948 | 1.000 | 1.000 | 1.000 | 1.000 | 1.000 |
| | | | 0.15 | 0.929 | 0.925 | 0.925 | 0.927 | 0.926 | 1.000 | 1.000 | 1.000 | 1.000 | 1.000 |
| | | | 0.30 | 0.896 | 0.892 | 0.891 | 0.894 | 0.893 | 0.999 | 0.999 | 0.999 | 0.999 | 0.999 |
| | | | 0.45 | 0.793 | 0.791 | 0.777 | 0.789 | 0.784 | 0.980 | 0.982 | 0.974 | 0.979 | 0.977 |
| | 200 | 200 | 0 | 0.987 | 0.987 | 0.987 | 0.987 | 0.987 | 1.000 | 1.000 | 1.000 | 1.000 | 1.000 |
| | | | 0.15 | 0.979 | 0.978 | 0.978 | 0.978 | 0.978 | 1.000 | 1.000 | 1.000 | 1.000 | 1.000 |
| | | | 0.30 | 0.960 | 0.958 | 0.957 | 0.959 | 0.958 | 1.000 | 1.000 | 1.000 | 1.000 | 1.000 |
| | | | 0.45 | 0.893 | 0.894 | 0.885 | 0.891 | 0.888 | 0.997 | 0.997 | 0.996 | 0.996 | 0.996 |
| | 50 | 100 | 0 | 0.687 | 0.606 | 0.687 | 0.687 | 0.670 | 0.987 | 0.982 | 0.989 | 0.988 | 0.988 |
| | | | 0.15 | 0.639 | 0.564 | 0.647 | 0.633 | 0.626 | 0.979 | 0.968 | 0.980 | 0.979 | 0.980 |
| | | | 0.30 | 0.586 | 0.508 | 0.587 | 0.579 | 0.571 | 0.950 | 0.931 | 0.950 | 0.950 | 0.949 |
| | | | 0.45 | 0.482 | 0.411 | 0.471 | 0.472 | 0.460 | 0.816 | 0.781 | 0.790 | 0.808 | 0.794 |
| | 100 | 200 | 0 | 0.929 | 0.904 | 0.929 | 0.929 | 0.925 | 1.000 | 1.000 | 1.000 | 1.000 | 1.000 |
| | | | 0.15 | 0.898 | 0.878 | 0.901 | 0.897 | 0.897 | 1.000 | 0.999 | 1.000 | 1.000 | 1.000 |
| | | | 0.30 | 0.858 | 0.831 | 0.860 | 0.856 | 0.855 | 1.000 | 0.999 | 1.000 | 1.000 | 1.000 |
| | | | 0.45 | 0.764 | 0.732 | 0.763 | 0.761 | 0.757 | 0.978 | 0.973 | 0.974 | 0.977 | 0.975 |

**Table 3** Average deviations from nominal 5% level of twelve tests using ANOVA for competing risks data

|  | TEST | Gaynor | | | | | Aalen | | | | | Pseudo | |
|---|---|---|---|---|---|---|---|---|---|---|---|---|---|
|  |  | Linear | Log | Llog | Arcs | Logit | Linear | Log | Llog | Arcs | Logit | Llog | Logit |
| $NUM_1$, | 50,50 | 0.496 | -0.929 | 0.009 | 0.301 | -0.261 | 0.238 | -1.054 | -0.256 | 0.089 | -0.456 | -0.186 | -0.013 |
| $NUM_2$ | 150,150 | 0.045 | -0.369 | -0.094 | 0.000 | -0.171 | -0.023 | -0.466 | -0.195 | -0.079 | -0.256 | -0.163 | -0.104 |
|  | 200,200 | 0.066 | -0.220 | -0.055 | 0.013 | -0.104 | -0.004 | -0.275 | -0.118 | -0.049 | -0.158 | -0.106 | -0.075 |
|  | 50,100 | 0.401 | -0.723 | -0.200 | 0.140 | -0.415 | 0.153 | -0.919 | -0.403 | -0.111 | -0.660 | -0.489 | -0.394 |
|  | 100,200 | 0.566 | -0.065 | 0.195 | 0.365 | 0.118 | 0.436 | -0.169 | 0.087 | 0.225 | -0.040 | 0.070 | 0.106 |
| TIME | 0.5 | 0.212 | -0.644 | -0.086 | 0.090 | -0.349 | 0.128 | -0.658 | -0.148 | 0.011 | -0.391 | -0.320 | -0.198 |
|  | 1.0 | 0.418 | -0.279 | 0.027 | 0.238 | 0.015 | 0.193 | -0.495 | -0.206 | 0.019 | -0.238 | -0.030 | 0.006 |
| CEN | 0.00 | 0.286 | -0.448 | 0.014 | 0.195 | -0.066 | 0.176 | -0.538 | -0.088 | 0.080 | -0.221 | -0.074 | 0.059 |
|  | 0.15 | 0.366 | -0.404 | 0.019 | 0.178 | -0.158 | 0.206 | -0.513 | -0.132 | 0.035 | -0.276 | -0.119 | -0.064 |
|  | 0.30 | 0.258 | -0.486 | -0.093 | 0.107 | -0.246 | 0.114 | -0.618 | -0.226 | -0.044 | -0.378 | -0.248 | -0.165 |
|  | 0.45 | 0.350 | -0.506 | -0.056 | 0.175 | -0.197 | 0.144 | -0.637 | -0.261 | -0.011 | -0.381 | -0.258 | -0.213 |
|  |  | 0.315 | -0.461 | -0.029 | 0.164 | -0.167 | 0.160 | -0.577 | -0.177 | 0.015 | -0.314 | -0.175 | -0.096 |

Upper panel: deviations given by $NUM_1\_NUM_2$ using Model 1; Middle panel: deviations given by *TIME* using Model 2; Lower panel: deviations given by *CEN* using Model 3; Last line: marginal effects of *TEST* from Model 4.

**Table 4** Average rejection rates for six tests using ANOVA for competing risks data

|  | TEST | Gaynor | | | | | Aalen | | | | | Pseudo | |
|---|---|---|---|---|---|---|---|---|---|---|---|---|---|
|  |  | Linear | Log | Llog | Arcs | Logit | Linear | Log | Llog | Arcs | Logit | Llog | Logit |
| $NUM_1$, | 50,50 | 63.645 | 60.939 | 61.454 | 62.886 | 61.780 | 61.952 | 59.139 | 58.103 | 60.826 | 58.838 | 62.124 | 62.901 |
| $NUM_2$ | 150,150 | 92.560 | 92.234 | 92.227 | 92.459 | 92.302 | 92.354 | 92.073 | 91.594 | 92.033 | 91.751 | 92.370 | 92.484 |
|  | 200,200 | 96.684 | 96.583 | 96.564 | 96.642 | 96.578 | 96.597 | 96.525 | 96.178 | 96.424 | 96.257 | 96.589 | 96.645 |
|  | 50,100 | 74.399 | 68.604 | 73.702 | 73.561 | 72.283 | 73.389 | 67.472 | 72.164 | 72.104 | 70.692 | 68.917 | 71.294 |
|  | 100,200 | 90.937 | 89.082 | 90.777 | 90.643 | 90.274 | 90.654 | 88.839 | 90.232 | 90.199 | 89.746 | 89.178 | 89.894 |
| TIME | 0.5 | 81.836 | 79.266 | 81.210 | 81.336 | 80.600 | 81.568 | 78.943 | 80.902 | 81.056 | 80.295 | 79.862 | 80.512 |
|  | 1.0 | 85.454 | 83.710 | 84.680 | 85.141 | 84.687 | 84.411 | 82.677 | 82.407 | 83.579 | 82.619 | 83.809 | 84.776 |
| CEN | 0.00 | 87.295 | 85.269 | 87.087 | 87.131 | 86.716 | 87.109 | 84.916 | 86.539 | 86.766 | 86.088 | 86.372 | 86.799 |
|  | 0.15 | 85.841 | 83.757 | 85.378 | 85.467 | 84.942 | 85.472 | 83.306 | 84.951 | 85.058 | 84.534 | 84.553 | 85.018 |
|  | 0.30 | 83.404 | 81.141 | 82.719 | 82.952 | 82.361 | 82.915 | 80.604 | 82.075 | 82.452 | 81.769 | 81.963 | 82.627 |
|  | 0.45 | 78.040 | 75.787 | 76.596 | 77.403 | 76.555 | 76.463 | 74.413 | 73.053 | 74.994 | 73.437 | 74.454 | 76.131 |
|  |  | 83.645 | 81.488 | 82.945 | 83.238 | 82.643 | 82.989 | 80.810 | 81.654 | 82.317 | 81.457 | 81.835 | 82.644 |

Upper panel: deviations given by $NUM_1\_NUM_2$ using Model 1; Middle panel: deviations given by TIME using Model 2; Lower panel: deviations given by CEN using Model 3; Last line: marginal effects of TEST from Model 4.

**Table 5** The comparing of ependymoma versus other brain tumors based on overall tests and tests at some fixed points in time

| Method | | At 1 years | At 3 years | At 5 years | At 7 years |
|---|---|---|---|---|---|
| Gaynor | Linear | **0.038 (4.320)** | 0.928 (0.008) | **0.040 (4.206)** | **0.025 (5.028)** |
| | Log | **0.040 (4.239)** | 0.928 (0.008) | 0.050 (3.827) | **0.032 (4.603)** |
| | Llog | **0.035 (4.431)** | 0.928 (0.008) | **0.043 (4.087)** | **0.028(4.812)** |
| | Arcs | **0.034 (4.508)** | 0.928 (0.008) | **0.042 (4.118)** | **0.027 (4.912)** |
| | Logit | **0.038 (4.318)** | 0.928 (0.008) | **0.045 (4.023)** | **0.029 (4.795)** |
| Aalen | Linear | **0.037 (4.364)** | 0.929 (0.008) | **0.045 (4.025)** | **0.028 (4.802)** |
| | Log | **0.035 (4.449)** | 0.929 (0.008) | 0.056 (3.649) | **0.037 (4.341)** |
| | Llog | **0.033 (4.555)** | 0.929 (0.008) | **0.048 (3.913)** | **0.032 (4.612)** |
| | Arcs | **0.032 (4.623)** | 0.929 (0.008) | **0.047 (3.938)** | **0.030 (4.686)** |
| | Logit | **0.034 (4.502)** | 0.929 (0.008) | 0.050 (3.846) | **0.033 (4.570)** |
| Pseudo | Llog | **0.030 (-2.164)** | 0.913 (0.109) | **0.040 (2.049)** | **0.024 (2.252)** |
| | Logit | **0.030 (-2.169)** | 0.913 (0.109) | **0.038 (2.073)** | **0.022 (2.282)** |
| Gray | | 0.174 (1.851)* | | | |
| Pepe and Mori | | 0.793 (0.069)* | | | |
| Renyi | | 0.479 (1.175)* | | | |
| KS | | 0.092 (0.211)* | | | |

The statistical results are expressed as "*P* values (Test statistic)"; Bold values are significance *P* values and statistical data; *: The statistical results for the overall CIFs.

**Table 6** The comparing of gender mismatch versus no gender mismatch based on overall tests and tests at some fixed points in time

| Method | | At 1000 days | At 2000 days | At 3000 days | At 4000 days | At 5000 days |
|---|---|---|---|---|---|---|
| Gaynor | Linear | 0.393 (0.730) | 0.248 (1.332) | 0.069 (3.307) | **0.039 (4.271)** | **<0.001 (11.610)** |
| | Log | 0.383 (0.761) | 0.236 (1.405) | 0.057 (3.612) | **0.028 (4.812)** | **<0.001 (17.679)** |
| | Llog | 0.390 (0.740) | 0.245 (1.354) | 0.066 (3.383) | **0.036 (4.388)** | **<0.001 (11.627)** |
| | Arcs | 0.389 (0.741) | 0.244 (1.357) | 0.065 (3.407) | **0.035 (4.439)** | **<0.001 (12.728)** |
| | Logit | 0.386 (0.752) | 0.240 (1.382) | 0.061 (3.503) | **0.032 (4.603)** | **<0.001 (13.925)** |
| Aalen | Linear | 0.393 (0.729) | 0.249 (1.329) | 0.069 (3.297) | **0.039 (4.243)** | **0.001 (11.117)** |
| | Log | 0.383 (0.760) | 0.270 (1.217) | 0.058 (3.601) | **0.029 (4.782)** | **<0.001 (16.995)** |
| | Llog | 0.390 (0.739) | 0.245 (1.351) | 0.066 (3.373) | **0.037 (4.359)** | **0.001 (11.136)** |
| | Arcs | 0.386 (0.751) | 0.240 (1.379) | 0.062 (3.493) | **0.032 (4.573)** | **<0.001 (13.354)** |
| | Logit | 0.390 (0.740) | 0.245 (1.354) | 0.197 (1.667) | **0.036 (4.409)** | **<0.001 (12.195)** |
| Pseudo | Llog | 0.384 (0.870) | 0.237 (1.183) | 0.059 (1.888) | **0.031 (2.161)** | **<0.001 (3.712)** |
| | Logit | 0.386 (0.868) | 0.238 (1.179) | 0.061 (1.874) | **0.032 (2.138)** | **<0.001 (3.510)** |
| Gray | | 0.064 (3.436)* | | | | |
| Pepe and Mori | | 0.216 (1.530)* | | | | |
| Renyi | | 0.143 (1.820)* | | | | |
| KS | | 0.014 (0.181)* | | | | |

The statistical results are expressed as "*P* values (Test statistic)"; Bold values are significance *P* values and statistical data; *: The statistical results for the overall CIFs.

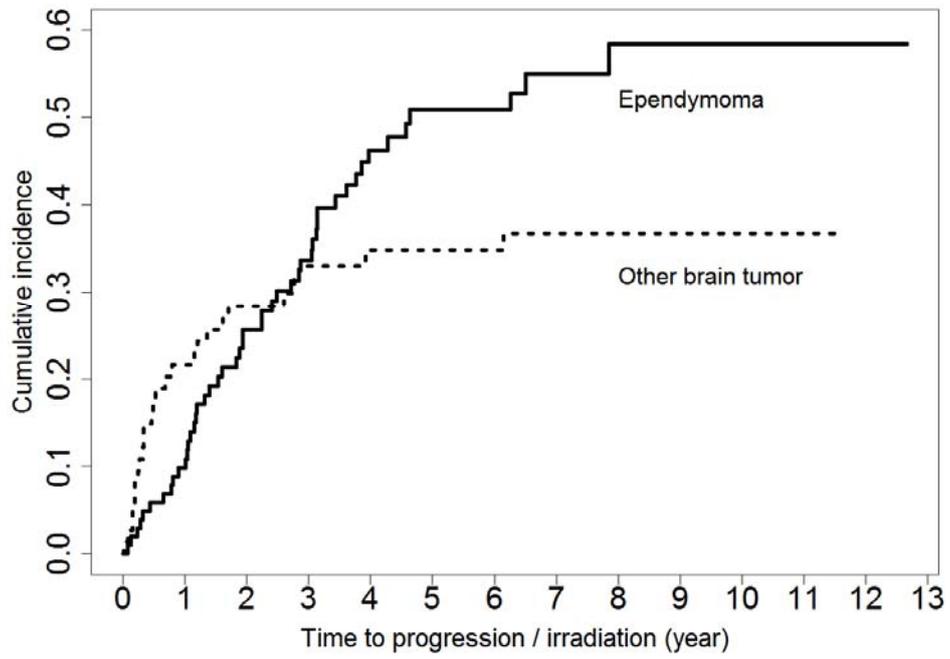

**Fig.1**  Cumulative incidence of progression/irradiation for ependymoma (solid line) and other brain tumor (dashed line) for the event irradiation after disease progression

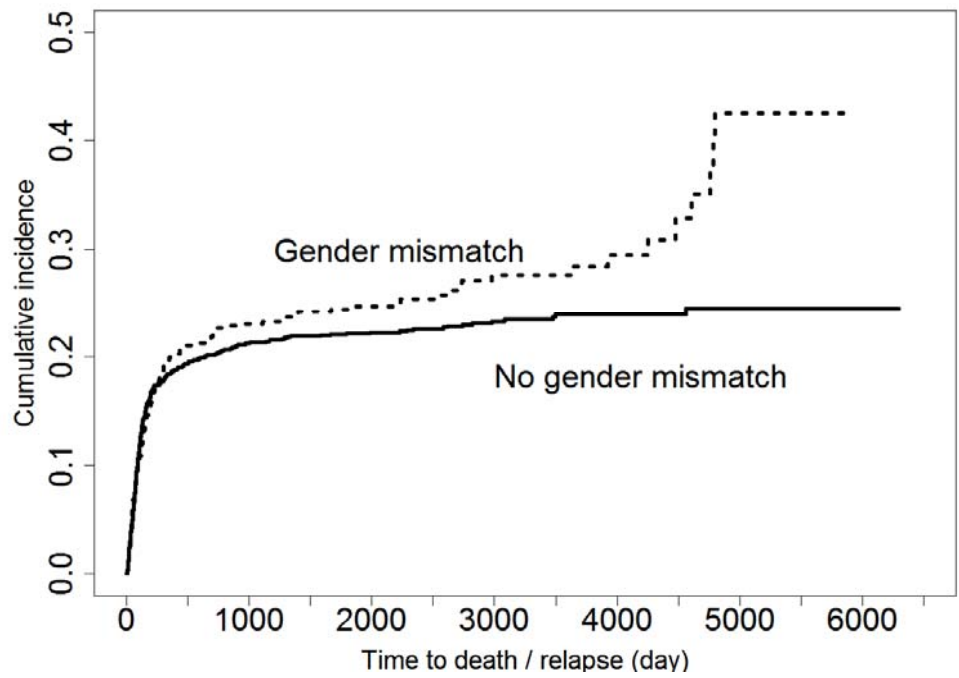

**Fig.2** Cumulative incidence of death/relapse for no gender mismatch (solid line) and gender mismatch (dashed line) for the event death from transplantation